\documentstyle[prl,multicol,aps,psfig]{revtex} 
\begin{document}
%
\draft
\title{ Phase Separation in Two-Dimensional Additive Mixtures}
\author{Arnaud Buhot and Werner Krauth
\footnote{buhot@physique.ens.fr; krauth@physique.ens.fr}}
\address{CNRS-Laboratoire de Physique Statistique,
Ecole Normale Sup\'{e}rieure,
24, rue Lhomond,\\ 75231 Paris Cedex 05, France}
\date{Received \today}
\maketitle
\begin{abstract}
We  study $2-$dimensional  binary mixtures of parallel squares as
well as of disks.  A recent cluster algorithm allows us to establish
an entropic demixing transition between a homogeneously packed
fluid phase and a demixed phase of a practically  close-packed
aggregate of large squares surrounded by a fluid of small squares.
\end{abstract}


\begin{multicols}{2}

\narrowtext 

Binary mixtures of impenetrable objects pose one of the important,
and lively, problems of statistical physics.  For many years it
has been discussed whether objects of different types $l$ (large)
and $s$ (small) would remain homogeneously mixed as the number of
these objects per unit volume increases. Particularly interesting
cases concern socalled additive mixtures~\cite{Asakura}, as hard
spheres with radii $r_l$ and $r_s$ or cubes with length $d_l$ and
$d_s$.  The problem of phase separation in binary mixtures is of
importance as the simplest model for colloids. It has been a meeting
ground for many different theoretical, computational, and experimental
approaches. As an example, the well-known closure approximations, as
well as virial expansions, both of very great importance for the
theory of simple liquids~\cite{Hansen}, have been brought to bear on this problem,
often with contradictory results.

\begin{figure}
\centerline{\psfig{figure=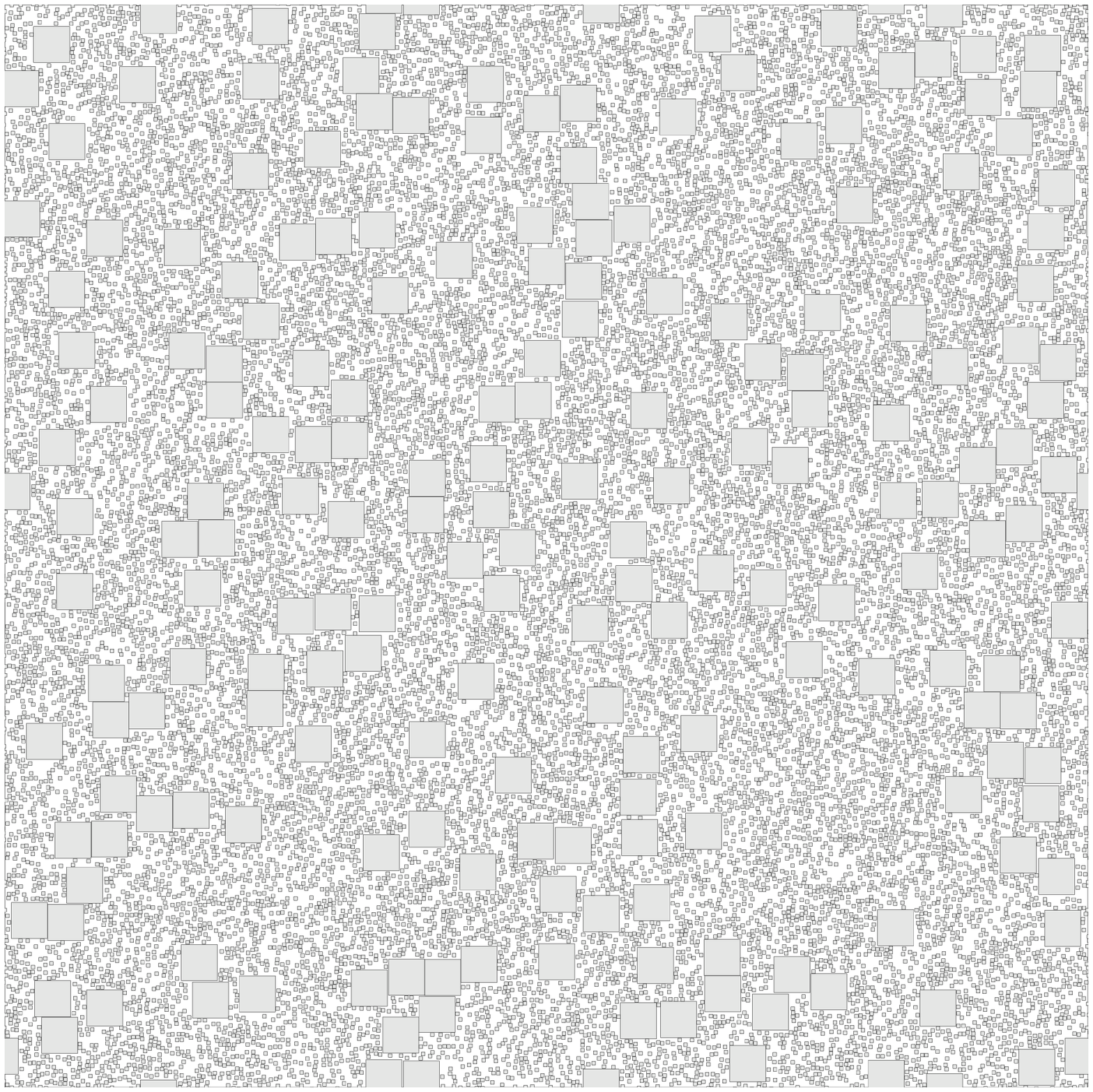,height=7cm} }
\caption{ Snapshot of $200$ big squares ($d_l=1$) and $20,000$ small
squares ($d_s=0.1$) in a periodically continued box of size $30 \times 30$
(packing fractions $\eta_l = \eta_s = 0.22$). 
}
\end{figure}

In this paper we discuss additive systems in two dimensions, parallel
hard squares and also hard disks.  For the parallel hard squares
system, we clearly see a phase separation transition which has not
been observed in previous work~\cite{Dijkstra,Cuesta}.  For hard
disks, we expect an analogous transition for extremely dissimilar sizes,
which must satisfy $r_s/r_l < 1/100$ for packing fractions $\eta_l
= \eta_s < 0.3$.

Monte Carlo simulations have long been performed on these systems.
They were recently boosted by a new cluster algorithm \cite{Dress,Buhot},
which allows thermalization of systems orders of magnitude larger
than previously possible.  The algorithm sidesteps a problem readily
apparent in Fig. 1, which shows a typical configuration in the
homogeneous phase. There, each large square is surrounded by very
many small objects. Trial moves of large squares will therefore be
rejected in the overwhelming majority of cases, and the algorithm
will get stuck quickly. Our algorithm rather swaps large patches
of the configuration in a way which preserves detailed balance
\cite{Dress,Buhot,Gould}. The algorithm is applicable for arbitrary
shapes and in any dimension, and in continuous  space as well as
on the lattice. Most importantly, the method works even for objects
very dissimilar in size as long as the total density is not too
high.

We are able to converge our simulations of two-dimensional parallel
squares for total packing fractions $\eta=\eta_s +\eta_l$ which do
not sensibly exceed the percolating density $\eta_{perc}$. As in
$3$D, we notice that the density $\eta_{perc}$ depends very little
on the ratio $R = d_s/d_l \leq 1$.  We find $\eta_{perc} \simeq
0.5$.  Figs 1 and 2 show snapshots of the simulations for $200$
large squares and $20,000$ small squares at equal composition
($\eta_s = \eta_l$) and total packing fractions $\eta = 0.44$ and
$\eta = 0.60$, respectively. It is evident that Fig. 1 represents
a homogeneous mixture whereas the system shown in Fig. 2 consists
of a `solid' block of large squares surrounded by a fluid of small
squares. In our opinion, these runs present direct evidence for a
transition of the homogeneously mixed fluid into a (solid-fluid)
phase.

\begin{figure}
\centerline{ \psfig{figure=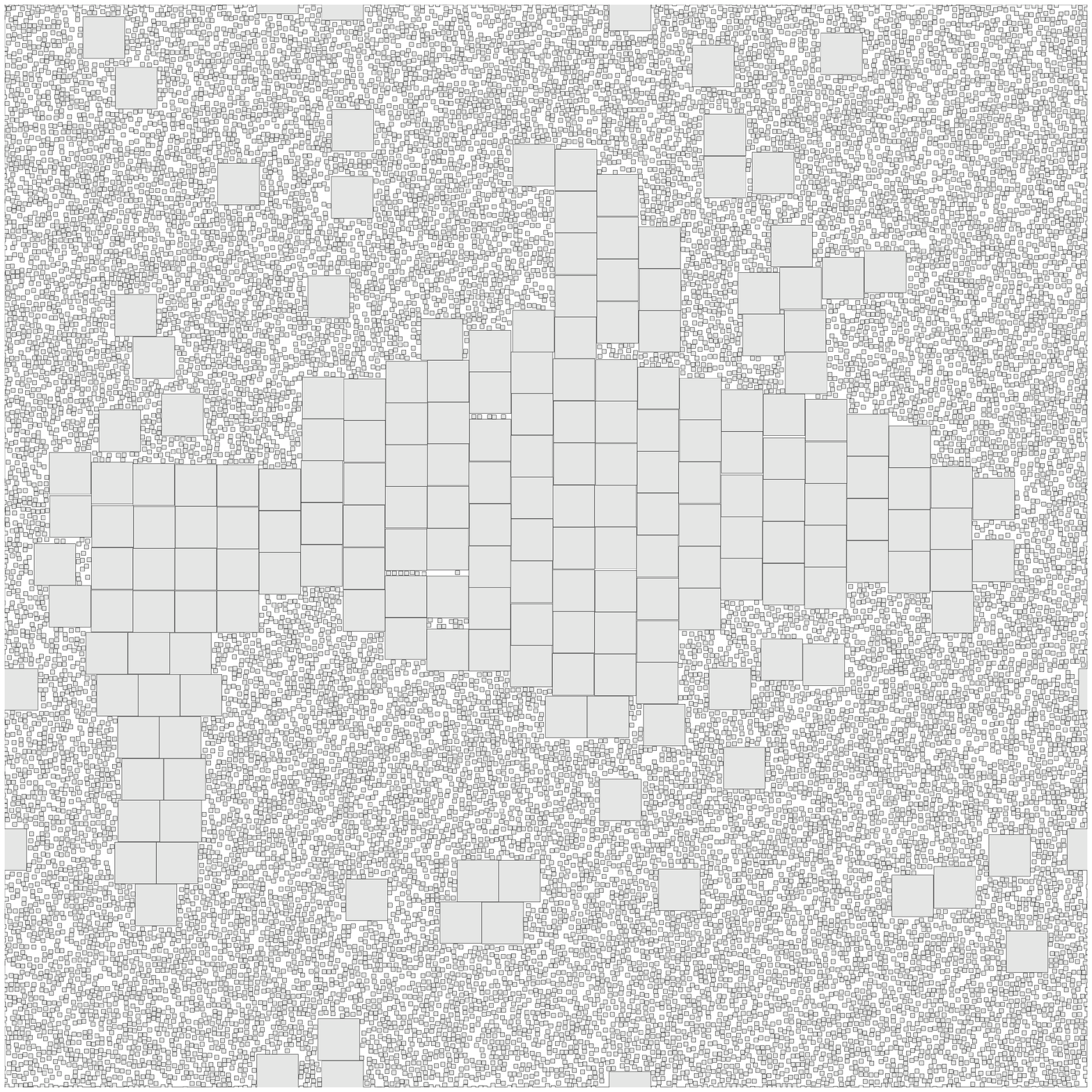,height=7cm} }
\caption{ Snapshot of $200$ big squares ($d_l=1$) and $20,000$ small
squares ($d_2=0.1$) in a periodically continued box of size $26 \times 26$
(packing fractions $\eta_l = \eta_s = 0.30$).}
\end{figure}

In the simulations at $\eta =0.6$, we have slightly exceeded the
percolation threshold $\eta_{perc}$. Therefore, the algorithm will
swap patches which usually comprise almost the whole system. This
generates problems for large systems, and we have {\em e.g.} been
unable to converge (at $\eta = 0.6$) a sample with $N_l = 800$,
$N_s=80,000$. In contrast, the simulations at lower packing fractions
converge extremely rapidly for arbitrary system size.  We can
summarize the situation for equal composition ($\eta_s = \eta_l$)
by the diagram of Fig. 3: The gray area corresponds to the region
of the diagram in which our algorithm performs extremely well. As
mentioned, this region is delimited for the homogeneous system by
the percolation threshold and by the appearance of high-density
areas as a consequence of phase separation.

We also studied the instability line as a function of the composition
$x = \eta_l/(\eta_s+\eta_l)$, especially since recent work of
Cuesta~\cite{Cuesta} suggests symmetry of the phase diagram with
respect to $x = 0.5$ in the $3$-dimensional system.  In the
$2$-dimensional system, simulations at different compositions ($x
= 0.3,0.5$ and $0.7$ for $R = 0.1$) contradict this behavior since
critical packing fractions are $\eta_{\rm crit} = 0.49 \pm 0.02$, $0.53 \pm
0.02$ and $0.60 \pm 0.05$, respectively.

In our three-dimensional simulation \cite{Buhot}, it was impossible
to interpret the data by direct inspection, as in Fig.  1 and Fig.
2.  We analyzed the transition therefore with the help of the
integrated pair correlation functions $G_{ll}(r) = 4 \pi \eta_l
\int r' dr' g_{ll}(r')$ \cite{Hansen}. We repeat the analysis in
the $2D$ case in order to stress the soundness of our procedure,
which considers $G_{ll}$ rather than the much noisier $g_{ll}$.
$G_{ll}(r) $ determines the average number of large particles around
a given large particle within a distance $r$. For the special case
of parallel hard squares, we define $r = \max(\Delta x, \Delta y)$,
where $\Delta x$ and $\Delta y$ are the two (periodically continued)
lateral distances.  With this definition, the distance of two
large squares in contact is $r = d_l$ and $G_{ll} =  \eta_l \int_{{\rm
max}(|x|,|y|)<r} g_{ll}(x, y) d  x d  y$ with $ g_{ll}$ the usual
pair correlation function \cite{Hansen}.  In Fig. 4, one can see
that the mixed system's $G_{ll}$ has pulled away from the pure
system's correlation function on all scales, a clear indication
that the large scale structure of the fluid has changed. The
staircase structure is a very clear indication of solid order.
Finally, the reason for the rather large finite-size effects at
large separation is immediately clear, since $G_{ll}(r)$ has to
meet the curve of the corresponding monodisperse system with $\eta
= \eta_l$ for half the box length.  A similar analysis was used to
establish the instability line in Fig. 3.

\begin{figure}
\centerline{ \psfig{figure=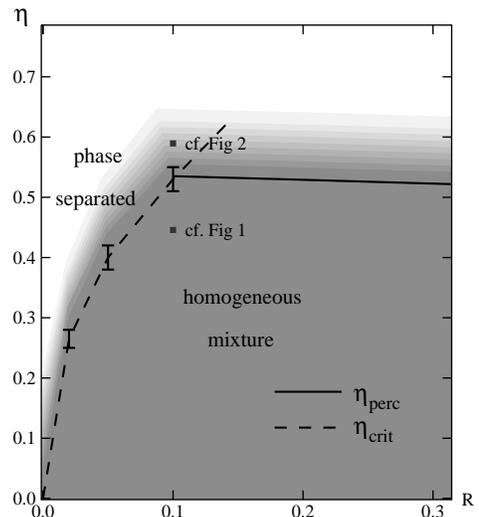,height=7cm} }
\caption{ Phases of the two-dimensional hard square system at equal
composition $\eta_s = \eta_l$ (total packing fraction $\eta = \eta_s 
+ \eta_l$ vs. $R = d_s/d_l$). The squares locate the
parameters of the snapshots in Figs 1 and 2. The region in which our
algorithm performs excellently is shaded in gray.
} 
\end{figure}

A comparison of Fig. 4 with the data for the homogeneous mixture
(cf. Fig. 1) at $\eta = 0.44$ is very reveiling. In the latter case
(not shown), the difference between the mixed and the monodisperse
system concerns mainly the region of small separation between the
squares and would be unobservable on the scale of Fig. 4.  The same
observation was made in $3$D.  The effects at small separations
$r$ are already detectable by visual inspection   of Fig. 1, since
there are quite many `bound' pairs and triplets. We find that each
large square has bound an average of $0.8$ squares.  This agrees
very nicely with our observation in $3D$, where we noticed the
onset of the phase-separation instability as the additional binding
was close to one.

We also performed simulations for mixtures of hard {\em disks}.
Let us begin our discussion with an indirect heuristic argument
which connects transition parameters for squares and for disks. It
has long been understood~\cite{Asakura} that the overlap of excluded
volumes entropically favors close contact of the large objects.
The excluded volume (for a small square) around a big square consists
in the area of the latter and a strip of width $d_s/2$ around it.
Side-to-side contact between two squares leads to an overlap of
excluded volume of the size $\Delta V_{\rm square} \sim d_l \times
d_s$.  As the large objects touch,  the volume available to the
small particles and therefore their entropy increase. At the same
time, contact of large objects {\em de}creases their contribution
to the entropy of the complete system.  The phase separation
transition appears when those two contributions are equals. As the
decrease of entropy due to contact of large objects may be considered
as independent of the ratio $R$ of the size of the small and large
objects, we may compare the increase of entropy due to larger
available volume for small disks or squares.  In fact, since the
small particles' contribution to the entropy is $S \sim N_s \log
V_s$ (with $V_s$ the available volume), one finds $\Delta S \sim
(N_s/V) \Delta V_{\rm square}$, and therefore 
\begin{equation} 
\Delta S_{\mbox{square}} \sim \eta_s d_l/d_s. 
\end{equation} 
Repeating the same calculation for disks, we notice of course that the 
overlap of excluded volume $ \Delta V_{\rm disk} \sim \sqrt{r_s^3 r_l}$ 
and the number of concerned small  disks are much smaller.  This leads to
\begin{equation} 
\Delta S_{\mbox{disk}} \sim \eta_s \sqrt{r_l/r_s}.
\end{equation} 
This order-of-magnitude argument tells us that (with
a hard square transition at $d_s/d_l \sim 1/10$ for $\eta \simeq 0.5$)
we can expect a transition for disks at best for $r_s/r_l \sim
1/100$.  Accordingly, our simulations for $r_s/r_l  \gtrsim 1/100$ 
at  $\eta = 0.6$ have reveiled no instability of the homogeneous
phase.  Even larger simulations, at $r_s/r_l= 1/150$, ($50$ large
and $1,250,000$ small disks, $\eta = 0.6$) did not converge, even
though the additional binding has continuously increased in the
course of a month-long simulation. In these simulations, the
depletion potential is very strong, but also extremely short-ranged.
The Monte Carlo simulation of such `golf-course' potentials is of
course extremely time-consuming and often impossible.

\begin{figure}
\centerline{ \psfig{figure=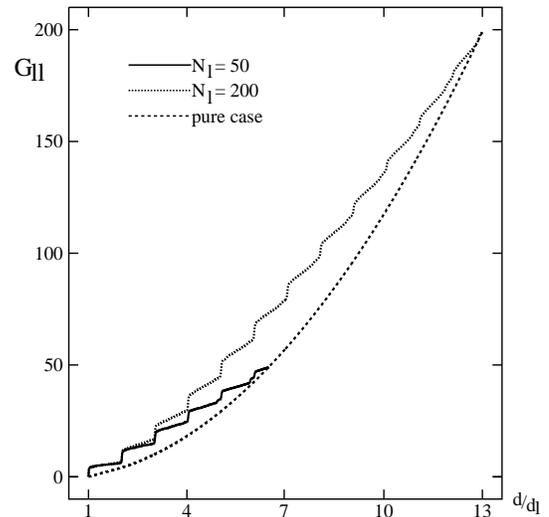,height=7cm} }
\caption{ Integrated two-point correlation function for the dense system
$\eta_l = \eta_s = 0.30$ (cf. Fig. 2) for $50$ and $200$ large squares,
respectively. The two curves are compared to the monodisperse system's
correlation function. 
}
\end{figure}

In conclusion we have studied the problem of phase separation of
two-dimensional systems (hard squares, hard disks) by direct Monte
Carlo simulation. For hard squares, our Monte Carlo data leave
little room to doubt a direct fluid to (solid-fluid) transition.
For hard disks, the stability of the homogeneous mixture seems
established for any `reasonable' ratio of radii $r_s/r_l \gtrsim 1/100$. 
Our heuristic argument would however lead us to expect a
transition for even more extreme ratios.

\end{multicols}
\end{document}